\newcommand{\DHA}[1]{{{#1}}}

\documentclass{aa}
\usepackage{txfonts}
\usepackage{graphicx}
\usepackage[dvipsnames]{xcolor}
\usepackage{hyperref}

\usepackage[utf8]{inputenc}
\usepackage{comment}

\begin{document}

\title{Inferring the origins of the pulsed $\gamma$-ray emission from the Crab pulsar with ten-year \emph{Fermi} LAT data}

\subtitle{}

\author{Paul K. H. Yeung\inst{1}}

\institute{Institute for Experimental Physics, Department of Physics, University of Hamburg, Luruper Chaussee 149, D-22761 Hamburg, Germany\\
\email{kin.hang.yeung@desy.de} }

   \date{Received April 14, 2020; accepted June 6, 2020}

\abstract
%context
{The Crab pulsar is a bright $\gamma$-ray source detected at photon energies up to $\sim$1~TeV. Its phase-averaged and phase-resolved $\gamma$-ray spectra below 10 GeV exhibit exponential cutoffs while those above 10~GeV apparently follow simple power-laws.}
%aims
{We re-visit the $\gamma$-ray properties of the Crab pulsar with 10-year \emph{Fermi} Large Area Telescope (LAT) data in the range of 60~MeV--500~GeV. 
With the phase-resolved spectra, we investigate the  origins and mechanisms responsible for the emissions.}
%methods
{The phaseograms are reconstructed for different energy bands and further analysed using a wavelet decomposition.
The phase-resolved energy spectra are combined with the observations of ground-based instruments like MAGIC and VERITAS to achieve a larger energy converage.
We fit power-law models to the overlapping energy spectra from 10~GeV to $\sim$1~TeV. We include in the fit a relative cross-calibration of energy scales between 
air-shower based gamma-ray telescopes with the orbital pair-production telescope of the Fermi mission.}
%results
{We confirm the energy-dependence of the $\gamma$-ray pulse shape, and equivalently, the phase-dependence of the spectral shape for the Crab pulsar. 
A relatively sharp  cutoff at a relatively high energy of $\sim$8~GeV is observed for the bridge-phase emission. 
The $E>$10~GeV spectrum observed for the second pulse peak is harder than those for other phases.}
{In view of the diversity of phase-resolved spectral shapes of the Crab pulsar, we tentatively propose a multi-origin scenario where the polar-cap, outer-gap and relativistic-wind regions are involved.}%conclusions


\maketitle


\section{Introduction} 
\label{sec:intro}
%% keyword of this paragraph: SN1054, 
The Crab pulsar and its nebula are products of the supernova explosion 
SN1054 and act as powerful particle accelerators. %ok
The Crab pulsar is one of the $239$  %introduce the exact number - quoting the FGL pulsar catalogue
pulsars whose $\gamma$-ray pulsations have been significantly
detected with the on-board \emph{Fermi} Large Area Telescope \citep[LAT;][]{Fermi_Fourth_2019}. 
Also, it is the only pulsar with pulsed emissions above 100 GeV robustly confirmed by the ground-based 
instruments MAGIC and VERITAS \citep[e.g.,][]{VERITAS_Crab_2011, Aleksic_Gap_2012}. Recently, pulsed emission
has been detected even up to TeV energies from the Crab pulsar \citep{ansoldi_teraelectronvolt_2016}.

The relevant emission mechanisms of $\gamma$-rays from pulsars are still under investigation. 
There have been a number of particle acceleration sites proposed as origins of pulsed $\gamma$-ray emission. %ok
The first one proposed is the polar cap region confined in the open magnetosphere at low altitudes 
\citep{Sturrock_PC_1971, Harding_PC_1978, Daugherty_PC_1982}. Due to rapid pair-creations under strong magnetic field, polar cap models predict a sharp super-exponential cutoff at several GeV which is 
not consistent with the observed $\gamma$-ray spectra of pulsars  \citep{abdo_second_2013}.

The second and third proposed regions are both located at high altitudes in the outer magnetosphere, 
respectively the slot gap
\citep[along the last open magnetic field lines;][]{Arons_SG_1983, Dyks_SG_2003, Muslimov_SG_2004} and outer gap
\citep[extending to the edge of the light cylinder;][]{Cheng_OG_1986, Romani_OG_1995, Cheng_OG_2000, Takata_OG_2006}. 
The \emph{Fermi} LAT pulse profiles and spectra of pulsars demonstrate that the responsible high-energy electron 
beams have a fan-like geometry scanning over a large fraction of the outer magnetosphere \citep{abdo_second_2013}. 
This favours the outer gap emission as a generally dominant component.

As observed with \emph{Fermi} LAT, MAGIC and VERITAS, at most on-pulse phases, the Crab pulsar's
spectrum above 10~GeV follows a rather hard power-law tail which extends beyond hundreds of GeV 
\citep{Aleksic_BD_2014, Nguyen_VERITAS_2015, ansoldi_teraelectronvolt_2016}. 
This certainly disfavours domination by the magnetospheric synchrotron-curvature mechanism, whose spectrum is theoretically expected to be well characterised by an exponential cutoff at several GeV due to magnetic pair-creations and/or radiation losses \citep{Cheng_OG_1986, Romani_SCR_1996, Muslimov_SG_2004, Takata_OG_2006, Tang_SCR_2008}. 
On the other hand, it is put forward by \citet{Harding_Gap_2015} that magnetospheric 
synchrotron-self-Compton (SSC) emission from leptonic pairs generated by cascades can account 
for the GeV spectral properties observed for the Crab pulsar. 

In addition, the fourth 
particle acceleration site, which is the relativistic wind located outside the light cylinder, is proposed as a responsible 
region as well \citep{Bogovalov_Wind_2000, Aharonian_Wind_2003, Aharonian_Wind_2012}. It is more recently suggested that the highest energy pulsed emission could be produced  in the region of the current sheet at a distance of 1--2 light cylinder radii \citep{Harding_Wind_2018} or even extending to tens of light cylinder radii \citep{Arka_Wind_2013, Mochol_Wind_2015}, where the kinetic-energy dominated wind is assumed to be launched.

It is noteworthy that the Crab pulsar has an energy-dependent $\gamma$-ray pulse shape, 
and equivalently, a phase-dependent spectral shape 
\citep[e.g.,][]{Fierro_EGRET_1998,  Abdo_Crab_2010, DeCesar_VelaP_2013}. 
This may suggest that emissions at different pulse phases are dominated by different emission regions.

In this work, we re-visit the $\gamma$-ray phaseograms and phase-resolved spectral energy distributions (SEDs) 
of the Crab pulsar, with the  $>$60~MeV LAT data accumulated over $\sim$10 years. 
Considering our LAT results in context with observations of ground-based instruments, 
we discuss the $\gamma$-ray origins for different phases individually.

\section{Data reduction \& analysis} \label{sec:data}
We perform a series of binned maximum-likelihood analyses (with an angular bin size of $0.1^\circ$) for a region of
interest (ROI) of $30^\circ\times30^\circ$  centered at RA=$05^\mathrm{h}34^\mathrm{m}31.94^\mathrm{s}$,
Dec=$+22^\circ00'52.2''$ (J2000), which is approximately the radio center of the Crab
Nebula \citep{lobanov_vlbi_2011}. 
We  use the data of 60~MeV--500~GeV photon energies,
registered with the LAT between 2008 August 4 and 2018 August 20. The data are reduced
and analyzed with the aid of the \emph{Fermi} Science Tools v11r5p3 package.

Considering that the Crab Nebula is quite close to the Galactic plane (with a
Galactic latitude of $-5.7844^\circ$), we adopt the events classified as
Pass8 ``Clean$"$ class for the analysis so as to better suppress the background.
The corresponding instrument response function (IRF) ``P8R2$_-$CLEAN$_-$V6$"$ is
used throughout the investigation.  We further filter the data by accepting
only the good time intervals where the ROI was observed at a zenith angle less
than 90$^\circ$ so as to reduce the contamination from the albedo of Earth.
\DHA{In phase-resolved analyses, we adopt the timing solution of the Crab pulsar provided by M. Kerr.}

In order to account for the contribution of diffuse background emission, 
we  include the Galactic 
background (gll$_-$iem$_-$v06.fits), the isotropic background \\
(iso$_-$P8R2$_-$CLEAN$_-$V6$_-$v06.txt) as well as all other  point sources
cataloged in the LAT 8-year Point Source Catalog \citep[4FGL;][]{Fermi_Fourth_2019}
within 32$^\circ$ from the ROI center in the source model.  We  set free the
spectral parameters of the  sources within 10$^\circ$ from the ROI center (including the prefactor and index of the Galactic 
diffuse background as well as the normalization of the isotropic background) in the
analysis. For the sources at angular separation beyond 10$^\circ$ from the ROI center, their
spectral parameters are fixed to the catalog values. 

The three point sources located within the nebula are cataloged 
as  4FGL J0534.5+2200, 4FGL J0534.5+2201i, and 4FGL J0534.5+2201s, which  model the
Crab pulsar, the IC, and synchrotron components of the Crab Nebula, respectively. 
In some cases, we fix the parameters of one or two components or even remove them from the source model, so as to avoid degeneracies in the fitting procedure.

\section{Results}

\subsection{LAT phaseograms at different energies}

First of all, we look into the LAT pulse profiles of the Crab pulsar in four energy bands: 60--600 MeV, 0.6--6 GeV, 6--60 GeV, and 20--500 GeV. We divide the full-phase into 50 bins (i.e., each bin covers a phase interval of 0.02). In the maximum-likelihood analysis for each bin, we remove 4FGL J0534.5+2201i and 4FGL J0534.5+2201s from the source model, and assign a single power-law (PL) to 4FGL J0534.5+2200 (i.e., the total emission of the Crab pulsar and its nebula is modelled as one component here). We adopt the same convention of phase as in \citet{buehler_gamma-ray_2012}.

\begin{figure*}
	\includegraphics[width=18cm]{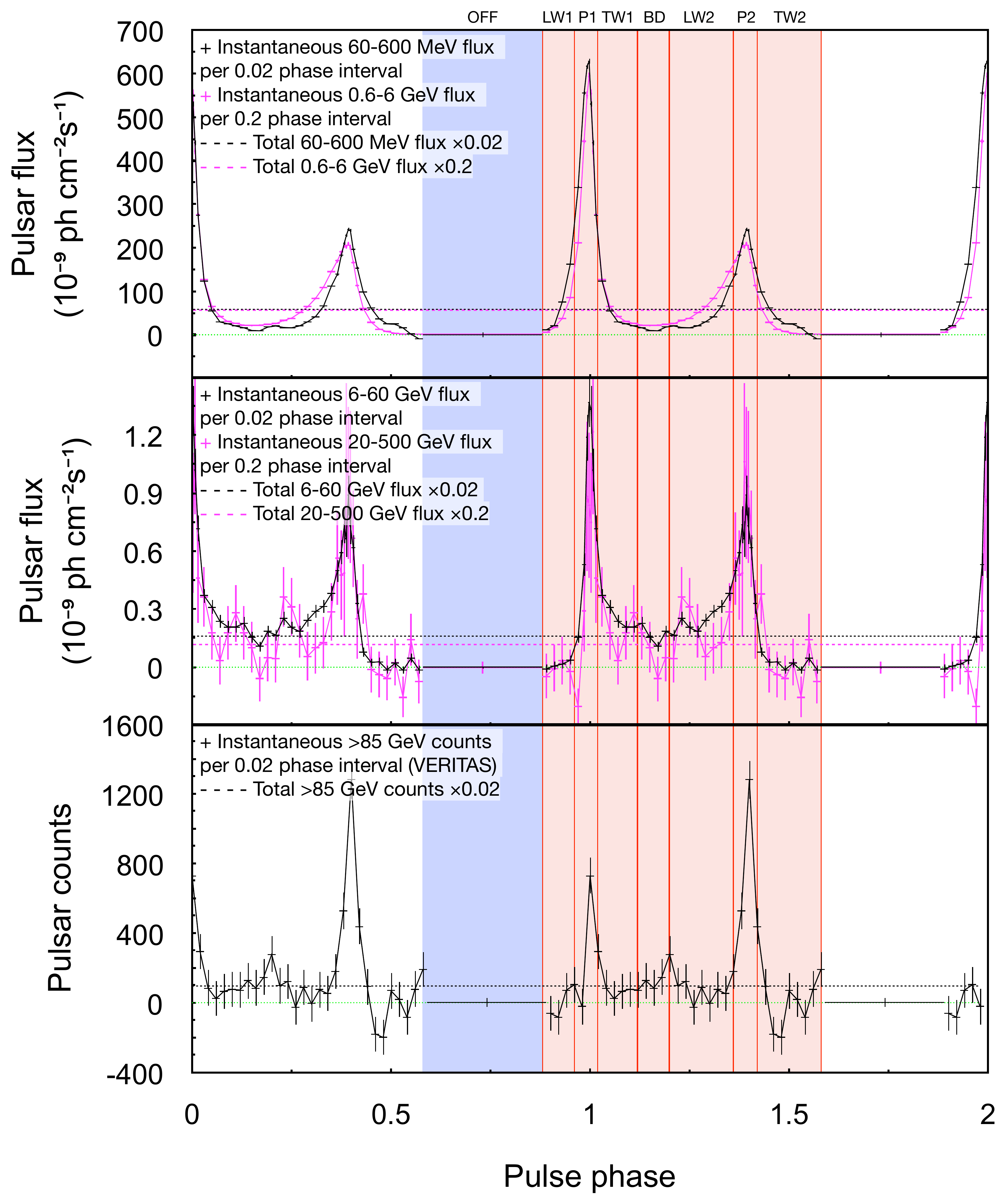}
	
	\caption{Phaseograms of the Crab pulsar in different energy bands. The >85 GeV phaseogram observed by VERITAS (the bottom panel) is taken from \citet{Nguyen_VERITAS_2015}. \label{phaseograms}}
\end{figure*}

The preliminary phaseograms show the first peak within the phase range of 0.98--0.02 and the second peak within phase 0.37--0.41. In order to localise the two peaks, we sub-divide phase 0.98--0.02 and 0.37--0.41 into bins of 0.01 phase interval, and then further sub-divide phase 0.99--0.01 and 0.38--0.40 into even smaller bins of 0.005 phase interval. Phase 0.58--0.88 is taken as the off-pulse region (thereafter OFF) where we determine the un-pulsed nebular fluxes. In each phaseogram, we combine those bins within OFF into 1 bin and then subtract the determined nebular flux from all bins, such that the flux within OFF is set at 0. The finalised phaseograms are plotted in Figure~\ref{phaseograms}, together with the >85 GeV phaseogram observed by VERITAS \citep{Nguyen_VERITAS_2015}. The phase-averaged flux in each energy band we investigate is also overlaid in the phaseogram.

Comparing all LAT phaseograms presented here, we have no evidence for any phase shift of the two peaks ($<1\sigma$; see Table~\ref{PeakPhases}). According to the pulse shapes, we divide the on-pulse region into 7 phase ranges for detailed analyses: Phase 0.88-0.96 (the leading wing of the first pulse; LW1), 0.96-0.02 (the peak region of the first pulse; P1), 0.02-0.12 (the trailing wing of the first pulse; TW1), 0.12-0.20 (the bridge between two pulses; BD), 0.20-0.36 (LW2), 0.36-0.42 (P2) and 0.42-0.58 (TW2).

\begin{table}
	\caption{Pulse peak phases determined from LAT phaseograms.}

\begin{tabular}{lcc}\hline\hline
Energy Band & Peak 1~$^\mathrm{a}$ & Peak 2~$^\mathrm{b}$   \\
(GeV) & &  \\ \hline
0.06--0.6  &  0.997 $\pm$ 0.004   &   0.393 $\pm$ 0.004  \\
0.6--6  &     0.998 $\pm$ 0.004   &   0.389 $\pm$ 0.005  \\
6--60  &      0.998 $\pm$ 0.004   &   0.391 $\pm$ 0.005  \\
20--500  &    0.003 $\pm$ 0.006   &   0.392 $\pm$ 0.004  \\ \hline
\end{tabular}

	\raggedright
$^\mathrm{a}$ It is calculated as the arithmetic mean of phases in 0.98--0.02 weighted by relative fluxes. \\
$^\mathrm{b}$ It is calculated as the arithmetic mean of phases in 0.36--0.42 weighted by relative fluxes.

	\label{PeakPhases}
\end{table}

In 60--600 MeV and 0.6--6 GeV, the ratios of maximum fluxes of P1 to P2 are $2.60\pm0.05$ and $2.85\pm0.08$, respectively. This ratio greatly decreases to $1.54\pm0.23$ in 6--60 GeV and to $1.14\pm0.59$ in 20--500 GeV. It further drops to an even smaller value of $0.57\pm0.09$ in the >85 GeV band.

From 60--600 MeV to 0.6--6 GeV, the fractional flux of BD increases from $(1.82\pm0.09)\%$ to $(3.15\pm0.07)\%$. This fraction further rises to $(8.38\pm0.73)\%$, $(4.7\pm4.4)\%$ and $(11.1\pm4.2)\%$ in 6--60 GeV, 20--500 GeV and >85 GeV respectively.

\subsection{Wavelet analyses on LAT phaseograms} \label{wlAnalyses}

\begin{figure*}
	\includegraphics[width=18cm]{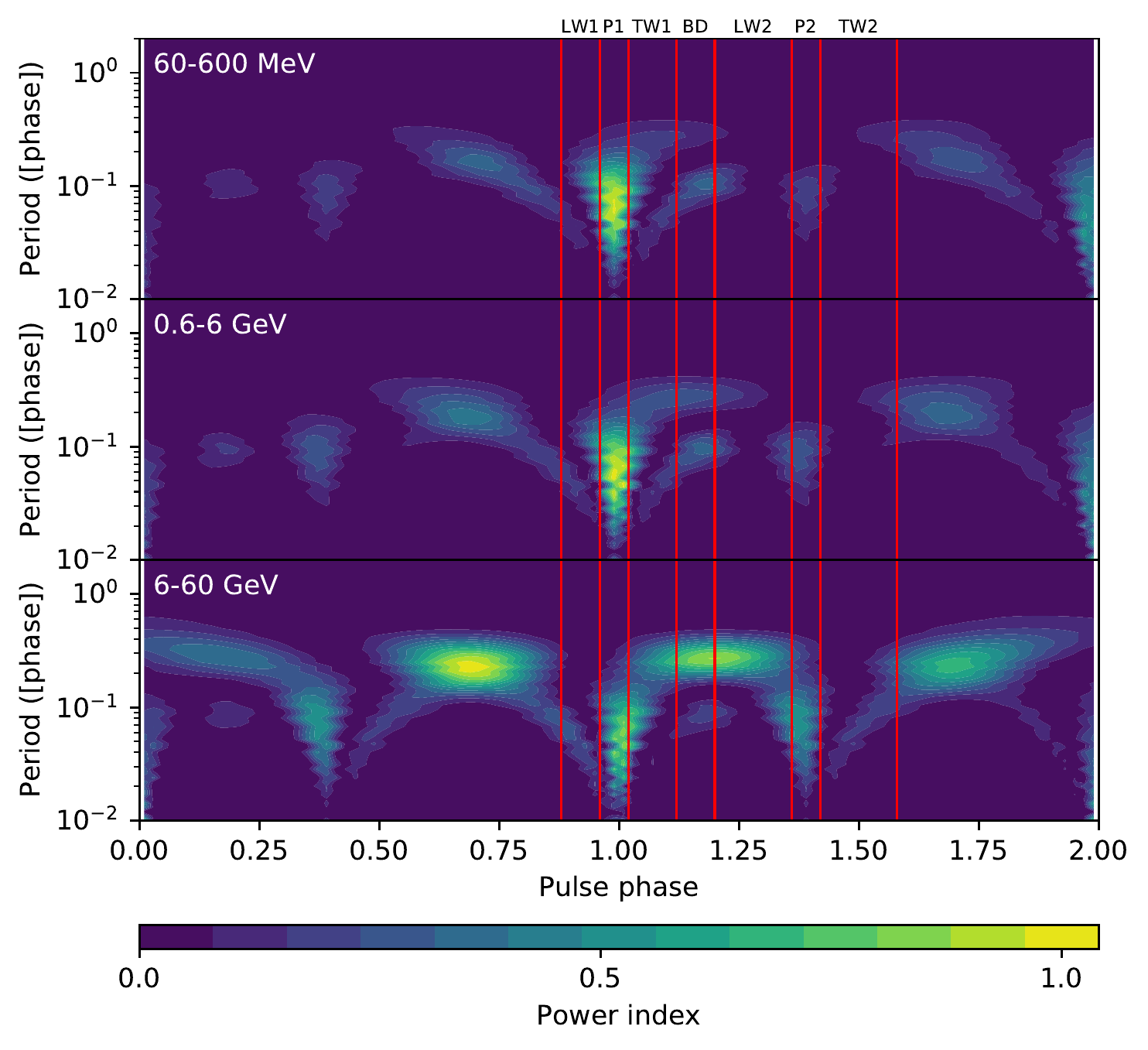}
	
	\caption{Wavelet maps of the preliminary LAT phaseograms in three exclusive energy bands. A continuous Fourier transform is applied in the period domain for each phaseogram. The Ricker wavelet is adopted. The power index (the colour scale) is defined as the variance per unit period per unit phase divided by the maximum. \DHA{Each red vertical line is a border between two phase ranges we define.} \label{wavelets}}
\end{figure*}

At any photon energy, the flux of the Crab pulsar changes exponentially during the wing phases. In order to investigate the instantaneous rates of flux change at different phases and energies, we apply continuous wavelet transforms to the preliminary LAT phaseograms which have a uniform bin size of a 0.02 phase interval. The Ricker wavelet is adopted. The photon statistics above 20~GeV are not sufficient for the wavelet transform. The results are shown in Figure~\ref{wavelets}.

The wavelet scale represents the timescale of flux change. Overall, at higher energies, the wavelet components at LW1 and TW2 extend to lower wavelet scales and become closer to vertical, while the components at TW1, LW2 and BD have greater power indices and become more tightly connected with each other.

Considering the pulse profiles and wavelet transformations synthetically, we derive a number of general trends: As the photon energy increases, \\
(a) the rate of flux increase in LW1 becomes faster, leading to a narrower wing;\\
(b) the rate of flux decrease in TW1 becomes slower, leading to a broader wing;\\
(c) the rate of flux increase in LW2 becomes slower, leading to a broader wing;\\
(d) the rate of flux decrease in TW2 becomes faster, leading to a narrower wing;\\
(e) the flux ratio of P1 to P2 drops;\\
(f) the fractional flux of BD rises.

The trends (a), (d) and (e) confirm what was reported in \citet{Abdo_Crab_2010}. In the following sub-sections, we further examine the trends (a)--(f) with spectral analyses.

\subsection{LAT SEDs for different pulse phases}

\subsubsection{Scheme of spectral analyses}

In broadband spectral analyses, we enable the energy dispersion correction which operates on the count spectra of most sources including the entire Crab pulsar/nebula complex, following the recommendations of the \emph{Fermi} Science Support Center. 

The energy spectrum of the un-pulsed nebular emission in the 60~MeV--100~GeV band is reconstructed by fitting a two-component (additive) model to the data collected during OFF. The flux normalisation of the pulsar component is fixed at 0. Similar to previous studies \citep{buehler_gamma-ray_2012, yeung_dimming_2019}, we assign the synchrotron component a PL with a photon index constrained within  3-5, and assign the IC component a log-parabola (LP):
\begin{center}
	$\frac{dN}{dE}=N_0(\frac{E}{10~\mathrm{GeV}})^{-(\alpha+\beta\mathrm{ln}(E/10~\mathrm{GeV}))}$ \ \ \ ,
\end{center}
where $\alpha$ is constrained within 0-2. It turns out that the synchrotron component has a PL index of $3.427\pm0.019$ and an integrated flux of $(2.500\pm0.018)\times10^{-6}$ ph cm$^{-2}$ s$^{-1}$ in the full phase, while the LP parameters of the IC component are determined to be $\alpha=1.759\pm0.023$, $\beta=0.106\pm0.014$ and $N_0=(5.12\pm0.14)\times10^{-13}$ cm$^{-2}$ s$^{-1}$ MeV$^{-1}$ (scaled to the full phase).

\begin{figure*}
	\includegraphics[width=18cm]{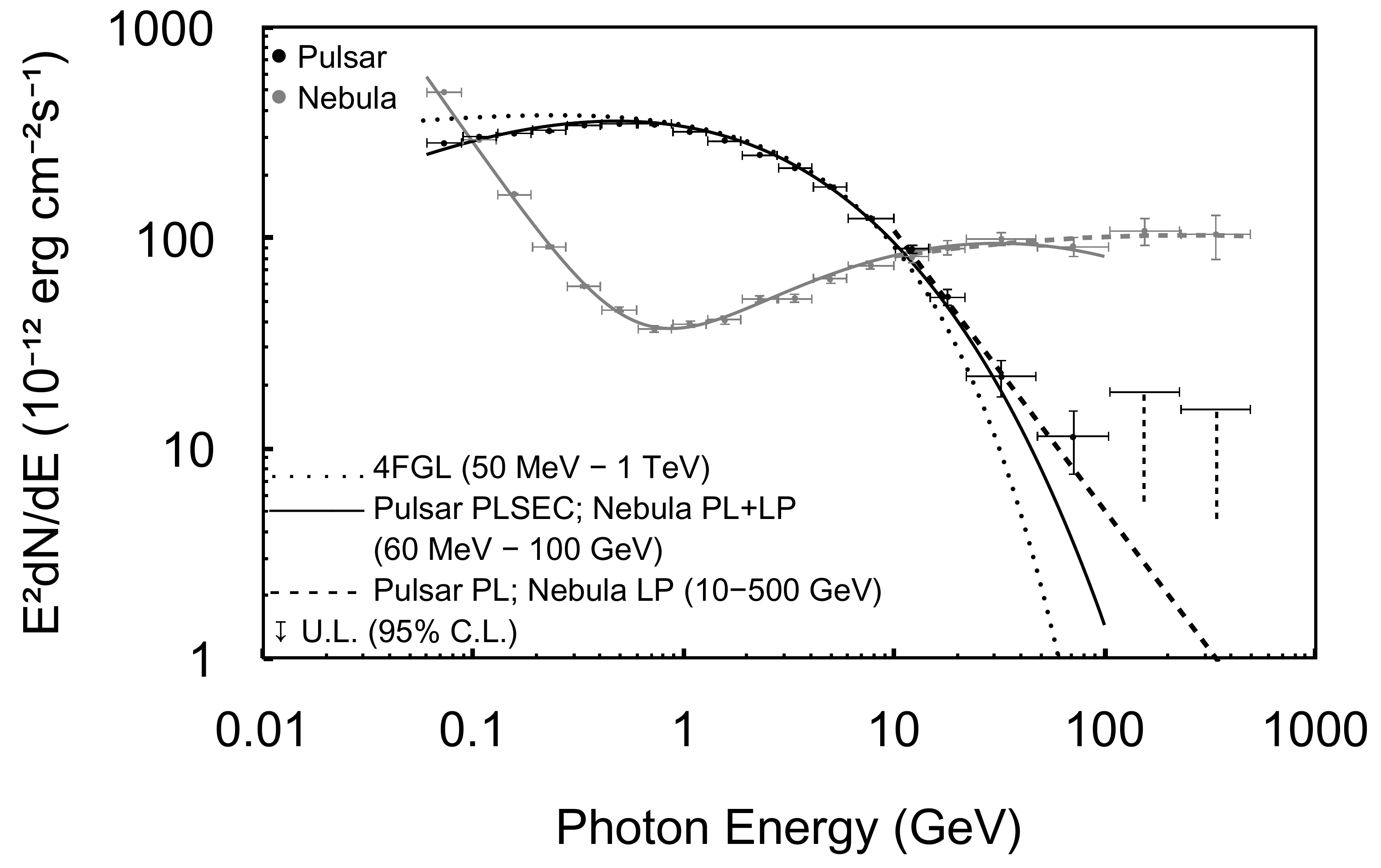}
	
	\caption{Phase-averaged LAT SEDs of the Crab pulsar and the Crab Nebula. The 4FGL model for the Crab pulsar, reconstructed with fixing $\lambda$ at 2/3, is overlaid for comparison. All upper limits presented are at a 95\% confidence level. We overlay the 60~MeV--100~GeV spectrum predicted by PLSEC and the 10--500~GeV spectrum predicted by PL.  \label{PhaAveSED}}
\end{figure*}

\begin{figure*}
	\includegraphics[width=18cm]{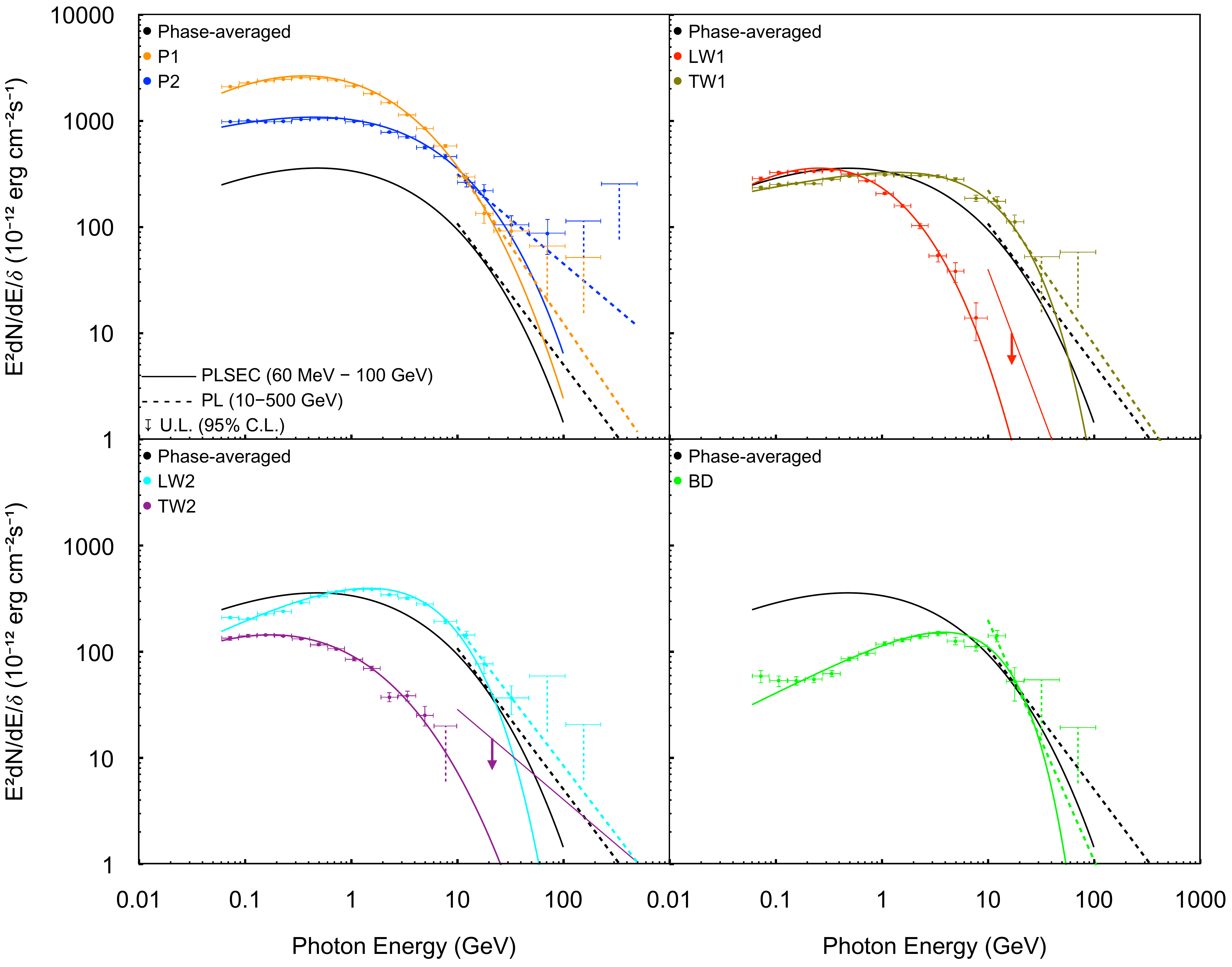}
	
	\caption{Phase-resolved LAT SEDs of the Crab pulsar. The vertical axis of each panel shows the differential flux per unit phase ($\delta$). All upper limits presented are at a 95\% confidence level. For each phase we investigate, we overlay the 60~MeV--100~GeV spectrum predicted by PLSEC and the 10--500~GeV spectrum predicted by PL. The model lines fit to the phase-averaged pulsar spectrum (Figure~\ref{PhaAveSED}) are also overlaid on each panel for comparison. Since the Crab pulsar is not significantly detected ($<2\sigma$) above 10~GeV at LW1 and TW2, the upper limits on differential fluxes at these energies and phases are represented by ``ad hoc" PL models \DHA{(the red and purple straight lines appended with arrows), each of which is determined through iterating the prefactor while fixing the index at the maximum-likelihood value.}  \label{LAT_SEDs}}
\end{figure*}

\begin{figure}
	\includegraphics[width=9cm]{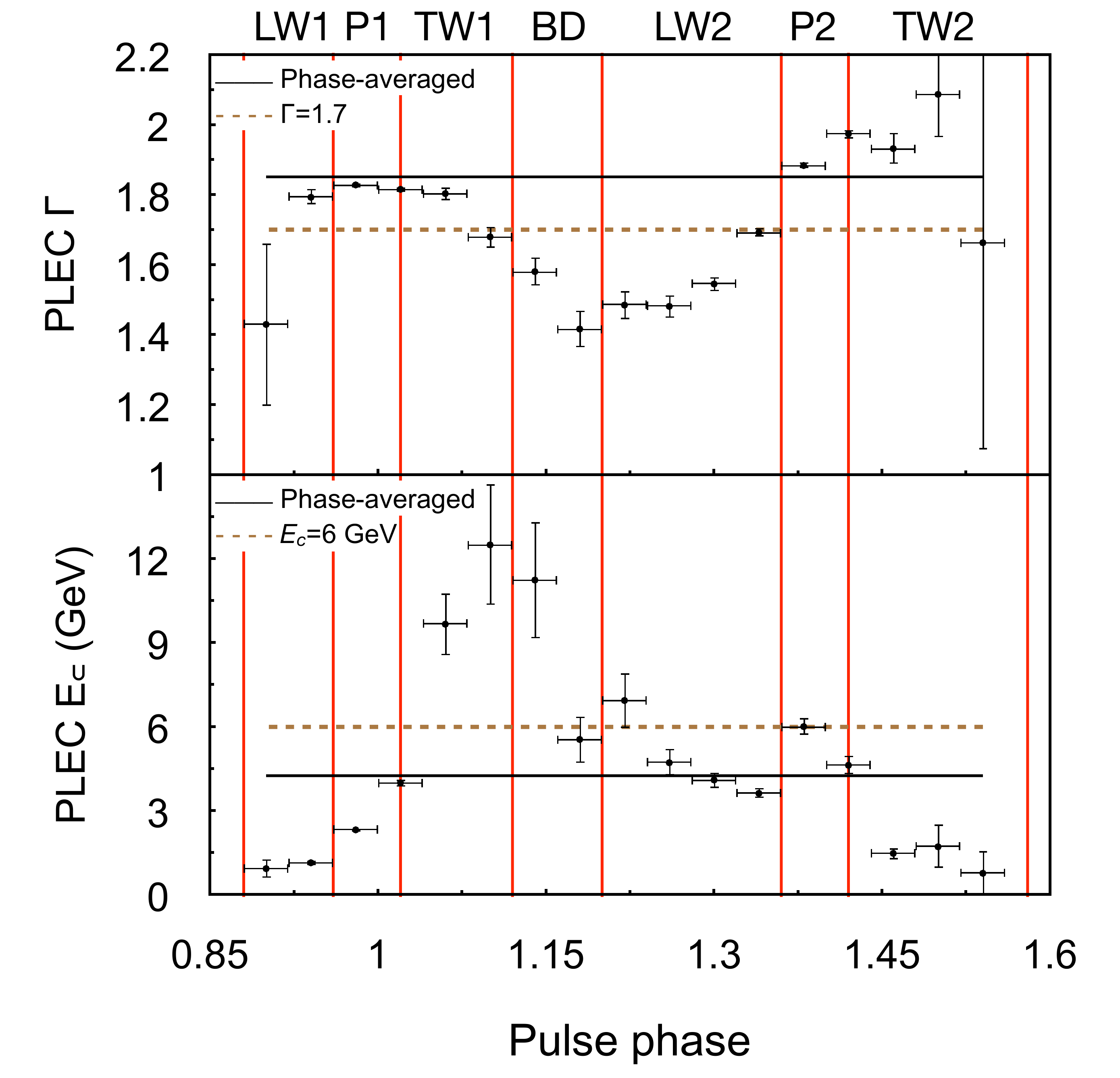}
	
	\caption{Photon indices $\Gamma$ and cutoff energies $E_c$ yielded by PLEC for individual phase bins of 0.04. \label{IndexGram}}
\end{figure}

\begin{figure}
	\includegraphics[width=9cm]{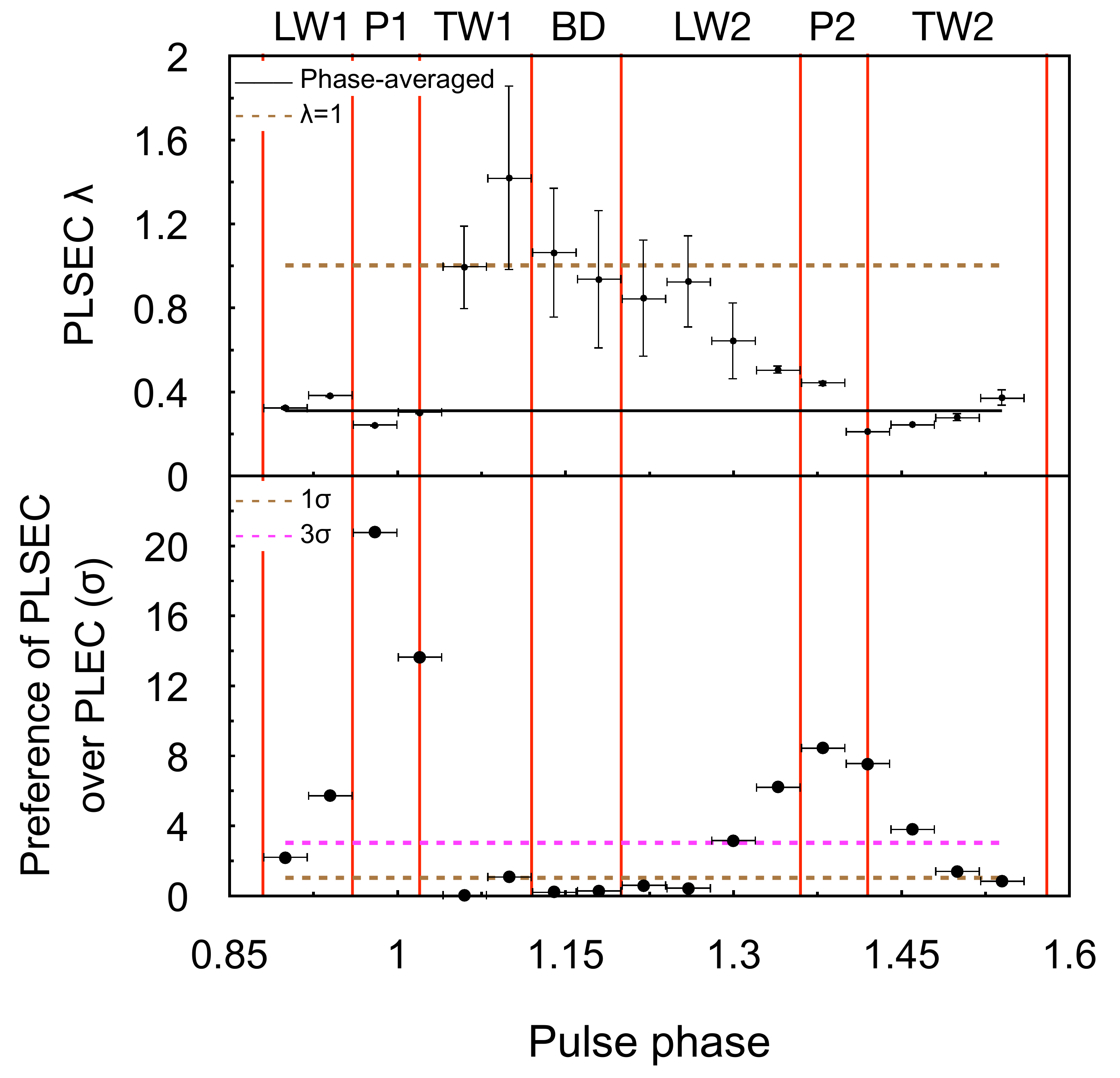}
	
	\caption{$\lambda$ values yielded by PLSEC and preference of PLSEC over PLEC for individual phase bins of 0.04. The latter is computed as the square root of the TS difference between those two models. \label{LambdaGram}}
\end{figure}

Then, we apply this nebular model to reconstruct the pulsar spectra at different phases in the same energy band. We examine how well the pulsar spectrum at each phase is described by, respectively, a power law with a super-/sub-exponential cutoff (PLSEC): 
\begin{center}
	$\frac{dN}{dE}=N_0(\frac{E}{E_0})^{-\Gamma}\mbox{exp}[-(\frac{E}{E_c})^{\lambda}]$ \ \ \ ,
\end{center}
and a power law with an exponential cutoff (PLEC) where $\lambda$ in PLSEC is fixed at 1. The pulsar parameters are left free while the nebular parameters are fixed at the determined values (with proper scalings to flux normalisations according to phase intervals). The obtained spectral models for the pulsar are tabulated in Table~\ref{60MeV}. The phase-averaged pulsar spectrum is plotted with the nebular spectrum in Figure~\ref{PhaAveSED}, and the phase-resolved pulsar spectra are plotted in Figure~\ref{LAT_SEDs}. Each presented flux has been scaled by the inverse of the phase interval (i.e., it refers to as the flux per unit phase). The spectral parameters for individual phase bins of 0.04 are plotted in Figures~\ref{IndexGram}~\&~\ref{LambdaGram}.

\begin{table*}
	\caption{60~MeV--100~GeV spectral properties of the Crab pulsar at different phases.}

\begin{tabular}{l|cc|cccc|c}\hline\hline
           &              \multicolumn{2}{c|}{\underline{PLEC}}               & \multicolumn{4}{c|}{\underline{PLSEC}}       &      \\
Phase      & $\Gamma$        & $E_c$        & $\Gamma$        & $E_c$       & $\lambda$            & $F$(60~MeV--100~GeV)~$^\mathrm{a}$       & $\Delta$TS~$^\mathrm{b}$  \\
           &              & (MeV)           &              & (MeV)          &              &            ($10^{-6}$~cm$^{-2}$~s$^{-1}$)             &      \\ \hline
Full-phase & 1.851 $\pm$ 0.004 & 4259   $\pm$ 80    & 1.328 $\pm$ 0.006 & 38.1   $\pm$ 1.5  & 0.308 $\pm$ 0.002 & 3.10              $\pm$ 0.01 & 512  \\ \hline
LW1        & 1.757 $\pm$ 0.021 & 1137   $\pm$ 55    & 0.888 $\pm$ 0.026 & 9.53   $\pm$ 0.76 & 0.351 $\pm$ 0.005 & 3.11              $\pm$ 0.03 & 32.1 \\
P1         & 1.827 $\pm$ 0.005 & 2750   $\pm$ 51    & 0.870 $\pm$ 0.007 & 0.51   $\pm$ 0.01 & 0.238 $\pm$ 0.001 & 22.93             $\pm$ 0.07 & 609  \\
TW1        & 1.831 $\pm$ 0.010 & 9181   $\pm$ 626   & 1.790 $\pm$ 0.033 & 7349   $\pm$ 1594 & 0.791 $\pm$ 0.122 & 2.64              $\pm$ 0.03 & 3.2  \\
BD         & 1.507 $\pm$ 0.032 & 7792   $\pm$ 922   & 1.501 $\pm$ 0.078 & 7607   $\pm$ 2522 & 0.979 $\pm$ 0.244 & 0.56              $\pm$ 0.03 & 0.01 \\
LW2        & 1.634 $\pm$ 0.010 & 4646   $\pm$ 175   & 1.500 $\pm$ 0.040 & 2244   $\pm$ 564  & 0.658 $\pm$ 0.062 & 2.38              $\pm$ 0.02 & 27.7 \\
P2         & 1.911 $\pm$ 0.006 & 6129   $\pm$ 259   & 1.660 $\pm$ 0.074 & 708    $\pm$ 612  & 0.415 $\pm$ 0.068 & 10.03             $\pm$ 0.06 & 93   \\
TW2        & 1.954 $\pm$ 0.026 & 1870   $\pm$ 164   & 1.029 $\pm$ 0.036 & 0.97   $\pm$ 0.09 & 0.256 $\pm$ 0.003 & 1.36              $\pm$ 0.02 & 17.7 \\ \hline
\end{tabular}

	\raggedright
$^\mathrm{a}$ The integrated flux per unit phase. \\
$^\mathrm{b}$ The difference in test-statistic (TS) between PLSEC and PLEC. Its square root is the significance at which PLSEC is preferred over PLEC. \\

	\label{60MeV}
\end{table*}

\begin{figure}
	\includegraphics[width=9cm]{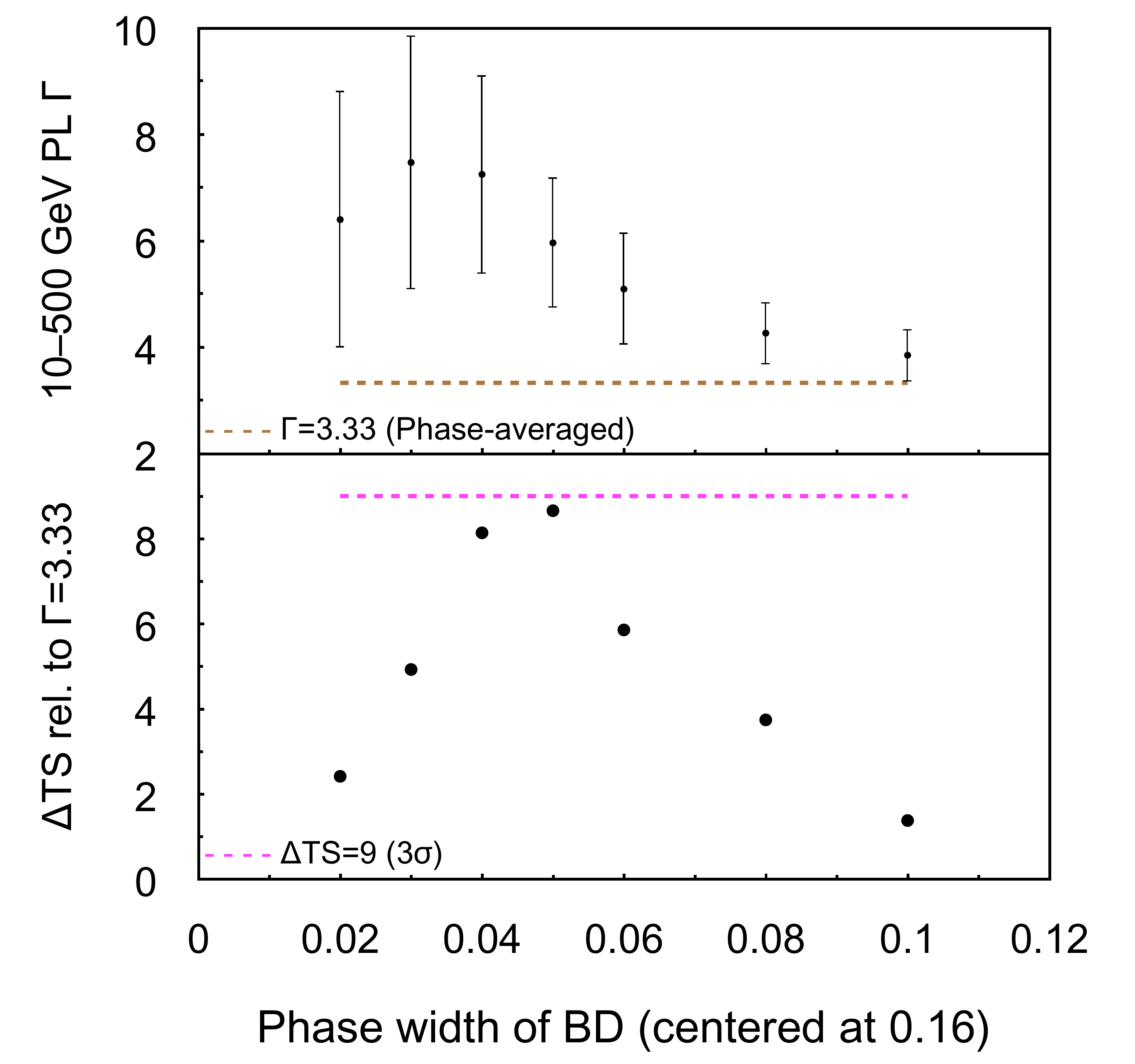}
	
	\caption{(top) Evolution of the 10--500~GeV PL index $\Gamma$ with adjustments to the phase width of BD. (bottom) Corresponding TS differences between fixing $\Gamma$ at the phase-averaged value of 3.33 and leaving it free. Each of their square roots is the significance at which the local spectrum for an adjusted phase interval of BD is softer than the phase-averaged spectrum. \label{EvolPhWidth}}
\end{figure}

We repeat this chain of exercise for the 10--500~GeV band. For the nebular emission, the synchrotron component is negligible, so we remove it from the source model. The nebular IC component is still modelled as a LP, while the pulsar component is modelled as a PL. The best-fit parameters for the nebula are $\alpha=1.86\pm0.14$, $\beta=0.023\pm0.047$ and $N_0=(5.10\pm0.42)\times10^{-13}$ cm$^{-2}$ s$^{-1}$ MeV$^{-1}$ (scaled to the full phase). The results are tabulated in Table~\ref{10GeV} and are overlaid in Figures~\ref{PhaAveSED}~\&~\ref{LAT_SEDs}. We also examine how significant the improvement is when we assign a curved model to the pulsar spectrum. Since the photon index at BD appears to be higher, we adjust the phase width of BD and study the evolution of the photon index (Figure~\ref{EvolPhWidth}). The narrowest phase width of BD we investigate is 0.05 for which the Crab pulsar is detected at a $\sim4\sigma$ significance.

\begin{table}
	\caption{10--500~GeV spectral properties of the Crab pulsar at different phases.}

\begin{tabular}{lccc}\hline\hline
Phase      & PL $\Gamma$       & $F$(10-500~GeV)~$^\mathrm{a}$         & TS    \\
           &                & ($10^{-9}$~cm$^{-2}$~s$^{-1}$)        &   \\ \hline
Full-phase & 3.33  $\pm$ 0.15 & 2.90            $\pm$ 0.14 & 698.9 \\ \hline
LW1        & 4.66  $\pm$ 0.63 & $<$0.69            & 0.4   \\
P1         & 3.47  $\pm$ 0.26 & 9.18            $\pm$ 0.79 & 319.7 \\
TW1        & 3.45  $\pm$ 0.29 & 5.68            $\pm$ 0.53 & 235.7 \\
BD         & 4.26  $\pm$ 0.59 & 3.82            $\pm$ 0.53 & 114.5 \\
LW2        & 3.31  $\pm$ 0.25 & 4.64            $\pm$ 0.40 & 256.8 \\
P2         & 2.84  $\pm$ 0.18 & 10.64           $\pm$ 0.84 & 354.4 \\
TW2        & 2.85  $\pm$ 1.06 & $<$0.97            & 4.0  \\ \hline
\end{tabular}

	\raggedright
$^\mathrm{a}$ The integrated flux per unit phase. \\

	\label{10GeV}
\end{table}

We proceed to generate binned spectra for the nebula and pulsar. We divide the 60~MeV--6~GeV band into 12 discrete energy bins (six bins per decade). 6--10 GeV is the 13th bin. The 10--500~GeV band is divided into five discrete bins, the first of which is further split into two. The procedures of broadband fittings are also applied to the spectral fittings of each bin. The nebular emission in each bin is modelled as a PL with an index fixed at a value derived from the broadband fitting. The results are overlaid in Figures~\ref{PhaAveSED}~\&~\ref{LAT_SEDs} as well. 

\begin{figure}
	\includegraphics[width=9cm]{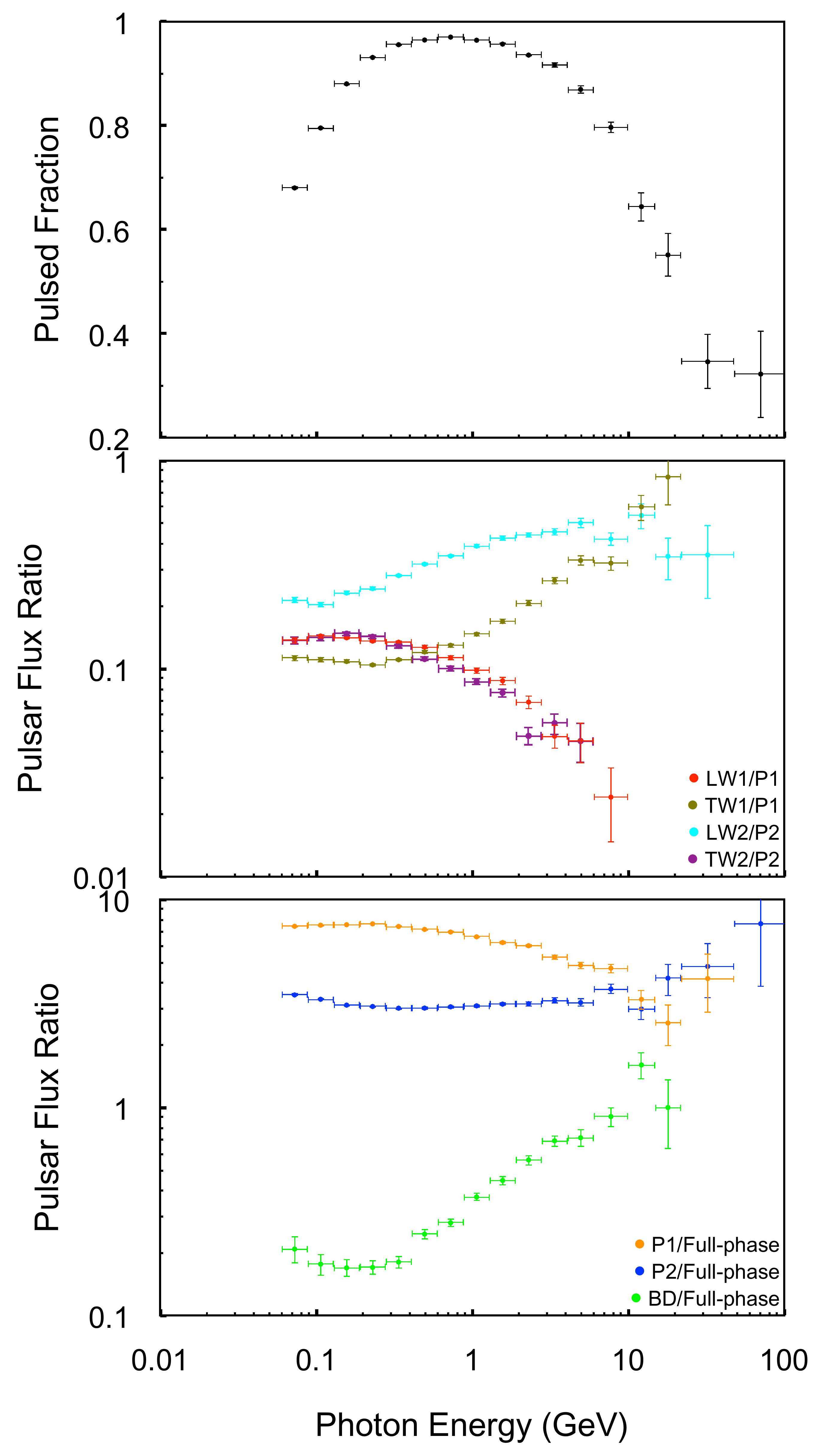}
	
	\caption{(top) Pulsed fraction of the entire Crab pulsar/nebula complex. (middle \& bottom) Flux ratios of the Crab pulsar between different pairs of phases in different energy bins. We recall that each flux has been scaled by the inverse of the phase interval. \label{flux_ratios}}
\end{figure}

Based on the binned spectra, we compute the pulsed fraction of the entire Crab pulsar/nebula complex, as well as the pulsar flux ratios between different pairs of phases at energies from 60 MeV to 100 GeV (plotted in Figure~\ref{flux_ratios}). The pulsed fraction is defined as $(F_{max}-F_{min})/(F_{max}+F_{min})$, where $F_{min}$ is the nebular flux, and $F_{max}$ is the pulsar flux in either P1 or P2 (the higher one) added to the nebular flux.

\subsubsection{Summary of spectral properties}

The pulsed fraction of the entire Crab pulsar/nebula complex is strongly dependent on the photon energy (see Figure~\ref{flux_ratios}). It is $\gtrsim90\%$ (and $\gtrsim80\%$) in 0.2--4 GeV (and 0.1--8 GeV). It drops to 25--50\% in 20--100 GeV. 

For the phase-averaged pulsar spectrum in 60~MeV--100~GeV, a PLSEC with $\lambda\approx0.31$ fits the data better than PLEC, indicating that a sub-exponential cutoff is strongly favoured. It is comforting to see that, in 0.2--14 GeV, this PLSEC model agrees within 15\% with the 4FGL model reconstructed with fixing $\lambda$ at 2/3.  The $\lambda$ values are widely varying with the pulse phase (see Figure~\ref{LambdaGram} and Table~\ref{60MeV}). Noticeably, during the phase 0.04--0.28, $\lambda$ is consistent with 1 within the tolerance of statistical uncertainties and PLSEC is not significantly preferred over PLEC ($\lesssim1\sigma$). In a PLSEC model, \DHA{there is a strong correlation of $\lambda$ with any other parameter}, making it nonsense to compare their values among different phases. Instead, we compare the $\Gamma$ and $E_c$ values of PLEC models among different phases. 

The PLEC model for the full-phase spectrum has a photon index $\Gamma\approx1.85$ and a cutoff energy $E_c=4.3\pm0.1$~GeV. During the phase 0.08--0.36 (and 0.16--0.28), PLEC fittings yield harder $\Gamma$ of $\lesssim$1.7 (and $\sim$1.45). During the phase 0.04--0.16 and 0.16--0.24, $E_c$ of PLEC is as high as $\sim$10.5~GeV and $\sim$6~GeV, respectively (see Figure~\ref{IndexGram}). These indicate that the total fractional flux of TW1, BD and LW2 generally increases with the photon energy. As follows, we summarise the $<$10~GeV spectral properties of the Crab pulsar based on PLEC and binned spectra (see Figures~\ref{PhaAveSED},~\ref{LAT_SEDs}~\&~\ref{flux_ratios} as well as Table~\ref{60MeV}), and relate them to the trends (a)--(f) derived in \S\ref{wlAnalyses}.
\begin{itemize}
	\item The $\Gamma$ value  at LW1 is lower than that at P1 by $0.07\pm0.02$. On the other hand, the $E_c$ value  at LW1 is about 40\% of that at P1. The flux ratio of LW1 to P1 is strongly decreasing in 0.5--10 GeV, confirming the trend (a). 
	\item $\Gamma$  at TW1 and that at P1 are consistent with each other within the tolerance of statistical uncertainties, and $E_c$  at TW1 is three times higher than that at P1. The flux ratio of TW1 to P1 is strongly increasing in 0.5--10 GeV, confirming the trend (b). 
	\item $\Gamma$  at LW2 is significantly lower than that at P2 by $0.28\pm0.01$. On the other hand, $E_c$  at LW2 is about three-fourth of that at P2. At $\sim$250 MeV, the flux ratio of LW2 to P2 starts rising with the photon energy significantly. This increment might come to an end at $>$4 GeV. Hence, we have strong evidence for the validity of the trend (c) in 0.25--4 GeV. 
	\item $\Gamma$  at TW2 is consistent with that at P2 (the difference is only at a $\sim1.6\sigma$ significance), and $E_c$  at TW2 is approximately 30\% of that at P2. At $\sim$350 MeV, the flux ratio of TW2 to P2 starts dropping with the photon energy significantly. This decrement might come to an end at $>$2 GeV. Hence, we have strong evidence for the validity of the trend (d) in 0.35--2 GeV. 
	\item Both the $\Gamma$ values  at P1 and P2 closely match the one for the full phase (the differences are $\lesssim$0.06). $E_c$  at P1 is about two-third of that for the full phase, which is about two-third of that at P2. In 60~MeV--1.5~GeV, the fractional flux of P1 is essentially uniform (the percent variance between any two bins is $\lesssim$20\%). Above 1.5~GeV, the fractional flux of P1 starts dropping with the photon energy. On the other hand, the fractional flux of P2 remains uniform at energies between 60~MeV and 10~GeV. Hence, the trend (e) is manifested in 1.5--10~GeV.
	\item BD has the lowest $\Gamma$ value  among all phases we investigate. It is much lower than $\Gamma$ for the full phase by $0.34\pm0.03$. Also, $E_c$  at BD is higher than that for the full phase by a factor of $\sim$1.8. The fractional flux of BD is robustly increasing in 0.35--10 GeV, confirming the trend (f). 
\end{itemize}
	
In 10--500~GeV, a PL is sufficient to describe the pulsar spectrum at each phase, and likelihood ratio tests indicate that any curved models are not statistically required ($\lesssim1\sigma$). Therefore, the energy-dependence of a flux ratio between two phases above 10 GeV can  be directly inferred from their difference in photon index (see Table~\ref{10GeV}). Exceptionally, the detection significance of the Crab pulsar at LW1 and TW2 is too low ($<2\sigma$) so that the ``ad hoc" PL models for these two phases cannot precisely predict the pulsar's differential flux. The photon indices of the Crab pulsar for  P1 and TW1 are  consistent with each other within the tolerance of statistical uncertainties. The indices for LW2 and P2  are  consistent with each other within the tolerance of $1.5\sigma$ uncertainties.  Therefore, we have no robust evidence for the validities of the trends (a)-(d)  at energies above 10 GeV. 

Interestingly, while the $<$3~GeV spectrum of the Crab pulsar is hardest at BD, its $>$10~GeV spectrum is apparently softest at BD. This is consistent with the relatively sharp  cutoff of the BD spectrum reported in Table~\ref{60MeV}. During the phase 0.135--0.185 (a central interval of BD), the fractional flux drops as $E^{-2.6\pm1.2}$  at a $\sim3\sigma$ significance (see Figure~\ref{EvolPhWidth}), indicating a potential reverse of the trend (f) above 10~GeV.

The validity of the trend (e) above 10 GeV is examined in the next sub-section with joint fits of the LAT and ground-based instruments' spectral points.

\subsection{Comparing LAT spectra to observations of ground-based instruments}

\begin{figure}
	\includegraphics[width=9cm]{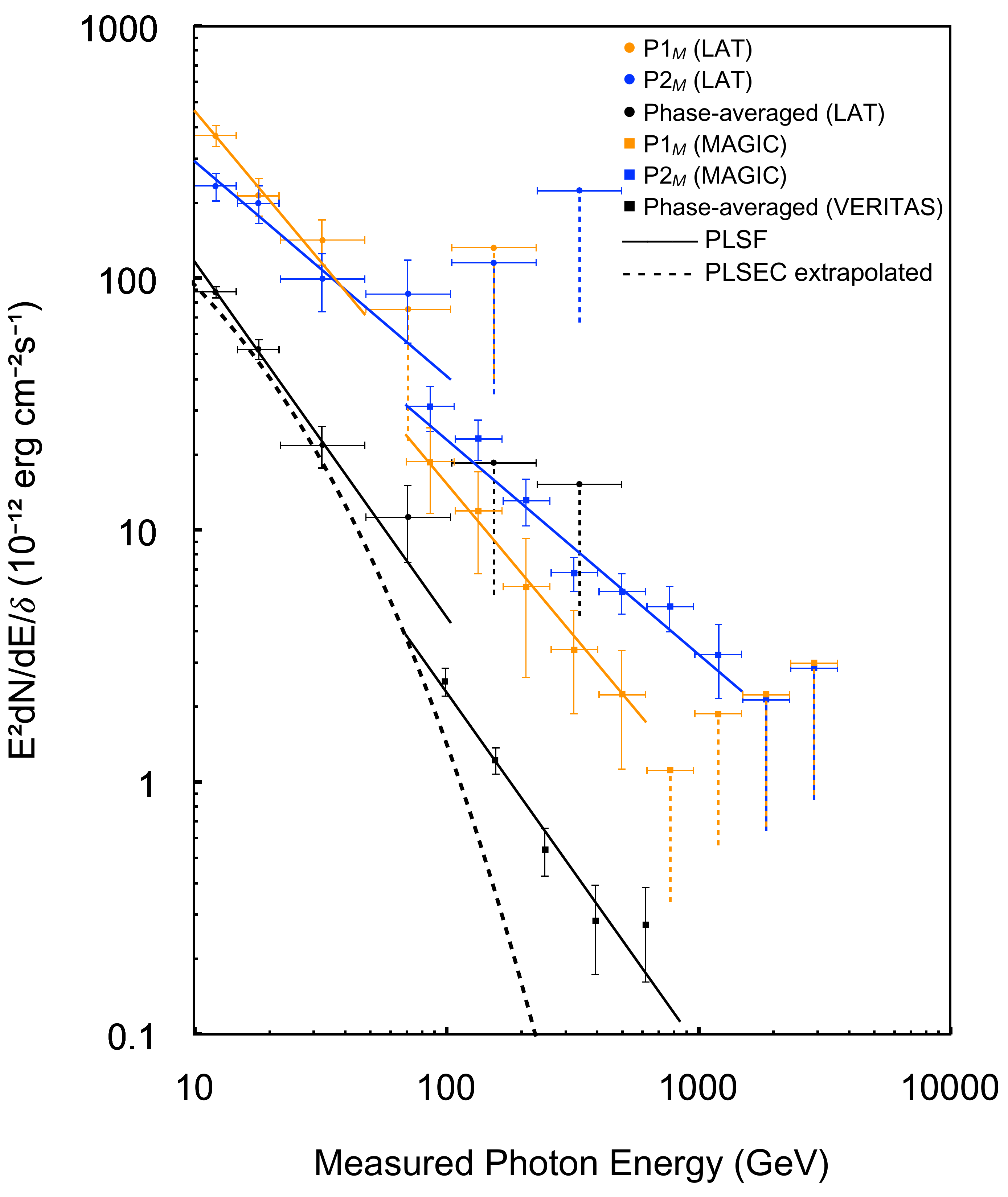}
	
	\caption{$>$10 GeV  SEDs of the Crab pulsar at different pulse phases observed with LAT and ground-based instruments. The vertical axis  shows the differential flux per unit phase ($\delta$). The horizontal axis shows the measured photon energy (unscaled). \DHA{Be noted that the  orange/blue LAT bins presented here are different from those presented in Figure~\ref{LAT_SEDs} because the phase ranges are adjusted for comparison purpose.} For each phase we investigate, we overlay the spectrum predicted by the joint-instrument fit of PLSF. The PLSEC fit to the broadband LAT spectrum in the full phase (Figure~\ref{PhaAveSED}) and its extrapolation are also overlaid for comparison. \label{GeV-TeV}}
\end{figure}

It is interesting to join the spectral data of LAT and ground-based instruments together for comparisons. Before that, we adjust the phase ranges of the two peaks to be the same as those defined in \citet{ansoldi_teraelectronvolt_2016} (Phase 0.983-0.026 and 0.377-0.422; thereafter P1$_M$ and P2$_M$ respectively). For these two phase ranges, we follow the same scheme to \DHA{re-compute} the LAT fluxes of the Crab pulsar in different energy bins starting from 10 GeV, which are plotted with the MAGIC fluxes in Figure~\ref{GeV-TeV}. For the full-phase, the LAT and VERITAS fluxes of the Crab pulsar at $>$10 GeV (the latter is taken from \citet{Nguyen_VERITAS_2015}) are overlaid in Figure~\ref{GeV-TeV} as well.

It is clearly demonstrated that the phase-averaged  VERITAS fluxes at energies above 80~GeV are higher than the extrapolated fluxes of the PLSEC fit to the broadband LAT spectrum. Also, for each phase interval we investigate here, \DHA{a PL is sufficient to describe the spectrum between 10~GeV and $\sim$1~TeV, and a spectral curvature or break is not statistically required ($<1\sigma$).}

\DHA{In order to take into account the differences in energy scale among LAT, MAGIC and VERITAS, we} fit the data set of each phase to a power law with a scaling factor on photon energies measured by ground-based instruments (PLSF):
\begin{center}
	$\frac{dN}{dE}=\begin{cases} N_0(\frac{E}{50~\mathrm{GeV}})^{-\Gamma} & \mbox{for LAT data} \\ N_0(\frac{\epsilon E}{50~\mathrm{GeV}})^{-\Gamma} & \mbox{for data of ground-based instruments} \end{cases}$ \ \ \ .
\end{center}

Since the data for P1$_M$ and P2$_M$ is collected by the same ground-based instrument, their data sets are fit together such that their solutions share the same scaling factor $\epsilon$. The results of fittings are presented in Table~\ref{joint_fittings} and the best-fit model lines are overlaid in Figure~\ref{GeV-TeV}.

\begin{table*}
	\caption{Joint fits of the LAT and ground-based instruments' spectral points for the Crab pulsar at different phases. The PLSF model is assumed. }

\begin{tabular}{l|c|ccc|c}\hline\hline
Phase      & Instruments    & $N_0$~$^\mathrm{a}$        & $\Gamma$       & $\epsilon$       & $\chi^2/d.o.f.$ \\
           &                & ($10^{-12}$~cm$^{-2}$~s$^{-1}$~GeV$^{-1}$)        &             &               &             \\ \hline
Full-phase & LAT \& VERITAS & 3.0  $\pm$ 0.5 & 3.41  $\pm$ 0.12 & 1.22    $\pm$ 0.12 & 3.1/6      \\ \hline
P1$_M$        & LAT \& MAGIC   & 16.9 $\pm$ 2.9 & 3.20  $\pm$ 0.13 & 1.23    $\pm$ 0.14 & 8.6/14     \\
P2$_M$        &                & 18.5 $\pm$ 2.6 & 2.85  $\pm$ 0.10 &                &            \\ \hline
\end{tabular}

	\raggedright
$^\mathrm{a}$ It has been scaled by the inverse of the phase interval. \\

	\label{joint_fittings}
\end{table*}

It is worth mentioning that the $\epsilon$ values obtained for MAGIC and VERITAS are both $\sim$1.22. Taking the statistical uncertainties into consideration, they are not significantly larger than 1 ($\le1.8\sigma$). It is also comforting to note that the best-fit $\epsilon-1$ values are only half of a fractional bin width of the ground-based instruments' data. Therefore, we have obtained physically reasonable fits.

It turns out that the photon index $\Gamma$ for P1$_M$ is lower than that for the full phase by only $\sim1.2\sigma$, while $\Gamma$ for P2$_M$ is lower than those for P1$_M$ and the full phase by $\sim2.1\sigma$ and $\sim3.6\sigma$ respectively. In other words, as the photon energy increases from 10~GeV to $\sim$1~TeV, the fractional flux of P1 remains constant or even slightly rises back, and that of P2 is significantly rising. The validity of the trend (e) is still suggested at energies $>$10 GeV. Since the   fluxes of LW1 and TW2 account for a negligibly small fraction of $\le$6\% (at a 95\% confidence level) above 10 GeV, the rise in total fractional flux of P1 and P2 implies a decline in total fractional flux of TW1, BD and LW2. This strengthens the interpretation that the trend (f) is reversed.

\section{Discussion and conclusion}

Our pulse profiles, wavelet transformations and spectral analyses for the Crab pulsar both demonstrate the strong dependence of the pulse shape on the photon energy, confirming previous studies. Equivalently, the LAT spectral shape of the Crab pulsar widely varies from phase to phase, indicating multiple origins of $\gamma$-ray emissions. According to the change in flux proportion among different phases with energy, the trends (a)--(f) derived in \S\ref{wlAnalyses} are generally valid below 10~GeV.

At any on-pulse phase we investigate, the broadband LAT spectrum of the Crab pulsar exhibits a (sub-)exponential cutoff at $E_c\lesssim10$~GeV. \DHA{We observe a higher PLEC cutoff energy at P2 than at P1  for the Crab pulsar. Interestingly, such a trend is predicted to occur for about 75\% of cases in the scenario of a dissipative  magnetosphere model}, where a relatively larger and azimuthally dependent electric field operates outside the light cylinder \citep{Brambilla_EField_2015}.  

In the framework of \citet{Lyutikov_OG_2012a} and \citet{Lyutikov_OG_2012b} for IC emission within the outer gap, the $\gamma$-ray spectrum of a pulsar could also manifest itself as a broken power law whose spectral break would correspond to a break in the electron distribution. This prediction is also consistent with a property observed by LAT and ground-based instruments: The Crab pulsar's spectrum from 10~GeV to $\sim$1~TeV follows a PL tail.

For the spectrum during the phase 0.04--0.24 (covering the whole BD), $\lambda$  yielded by PLSEC is \DHA{$\gtrsim$1}  and $E_c$ yielded by PLEC is $\sim$(6--10.5)~GeV, consistent with an inherent feature \DHA{(a super-exponential cutoff)}  predicted in polar cap models \citep[e.g.,][]{deJager_PolarCap_2002, Dyks_PolarCap_2003}. Such a relatively sharp  cutoff is explainable in terms of strong magnetic absorption of low-altitude $\gamma$-ray photons above 10 GeV. However, the bridge emission of the Crab pulsar is significantly detected by MAGIC at energies up to 200~GeV \citep{Aleksic_BD_2014}. The PL indices of its $>$10~GeV LAT spectrum (in this work) and $>$50~GeV MAGIC spectrum \citep{Aleksic_BD_2014} are both $\sim$4. This is not expected in polar cap models.

During LW1, P1, P2 and TW2, a sub-exponential cutoff with $\lambda\le0.6$ is strongly favoured to describe the observed spectrum. This rules out the polar cap origin and suggests high-altitude emission zones for these phases. While \DHA{a traditional} outer gap model naturally explains the sub-exponential cutoff detected for the Geminga pulsar \citep{Ahnen_Geminga_2016}, it can only account for \DHA{the emissions of the Crab pulsar  at energies no higher than 10~GeV.} The  Crab pulsar's spectrum at each pulse peak exhibits a $>$10~GeV PL tail (in this work) which is much harder than that of the Geminga pulsar \citep{Ahnen_Geminga_2016}, and the tail for P2 even extends to 1.5~TeV without a spectral break \citep{ansoldi_teraelectronvolt_2016}. Therefore,  a more complicated scenario  is required to explain the spectral properties of the Crab pulsar at P1 and P2. 

The IC $\gamma$-rays (including SSC emission) from magnetospheric acceleration gaps, with ``ad hoc" modifications to the models, can roughly match the P1 and P2 fluxes of the Crab pulsar at energies up to 400 GeV \citep[e.g.,][]{Aleksic_Gap_2011, Aleksic_Gap_2012, Harding_Gap_2015, Osmanov_magnetocentrifugal_2016}. Impressively, \citet{Harding_Gap_2015} took into account the primarily accelerated electrons as well as the leptonic pairs generated by cascades, \DHA{and \citet{Osmanov_magnetocentrifugal_2016} considered magnetocentrifugal particle acceleration which is efficient close to the light cylinder}.  Proposed feasible alternatives include wind models, where pulsed $\gamma$-rays are due to synchrotron and/or IC radiation from relativistic electrons outside the light cylinder. \citet{Aharonian_Wind_2012} modelled the pulsed $\gamma$-ray emission of the instantaneously accelerated wind, while \citet{Arka_Wind_2013} and \citet{Mochol_Wind_2015} modelled that of the wind current sheet.

The trends (a) and (d) of the Crab pulsar entail a decrease in pulse width with increasing energy, which is also detected for the Vela pulsar \citep{DeCesar_VelaP_2013, HESS_VelaP_2018}. Such a phenomenon is naturally explained by wind models as well. Besides, a harder spectrum at P2 compared to P1 (i.e., the trend (e)) is observed for both Crab and Vela pulsars \citep[this work;][]{DeCesar_VelaP_2013, HESS_VelaP_2018}. This could be attributed to an anisotropy of the wind \citep{Aharonian_Wind_2012}. Furthermore, wind models predict the bridge emission above 50 GeV as well, but a more complicated density profile of the wind is required to reproduce the observed flux proportion among the bridge and two peaks  \citep{Aharonian_Wind_2012, Khangulyan_Wind_2012}.

\begin{figure}
	\includegraphics[width=9cm]{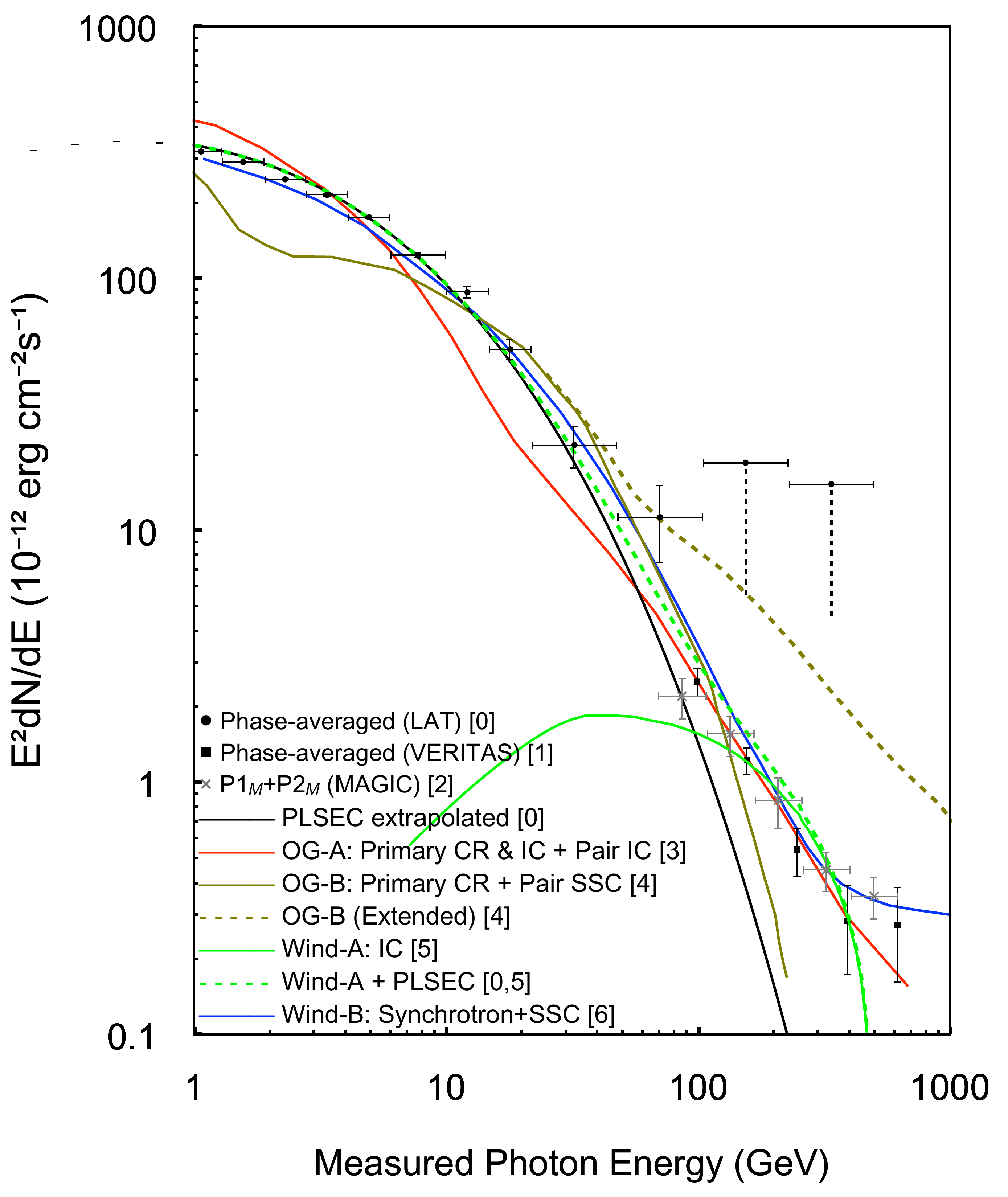}
	
	\caption{Comparison of the observed full-phase spectrum with four theoretical models, namely OG-A, OG-B, Wind-A and Wind-B. Descriptions of points and lines are at the bottom-left corner. OG stands for ``outer gap". CR stands for ``curvature radiation". ``Primary" means emission from primarily accelerated electrons. ``Pair" means emission from leptonic pairs generated by cascades. ``OG-B (Extended)" is the nominal OG-B model modified with a power-law extension to the cascade pair spectrum. References: [0] this work, [1] \citet{Nguyen_VERITAS_2015}, [2] \citet{ansoldi_teraelectronvolt_2016}, [3] \citet{Aleksic_Gap_2012}, [4] \citet{Harding_Gap_2015}, [5] \citet{Aharonian_Wind_2012}, [6] \citet{Mochol_Wind_2015}.  \label{ShowCase}}
\end{figure}

For the phase-averaged spectrum in GeV--TeV, a schematic comparison of observational results with different theoretical predictions is shown in Figure~\ref{ShowCase}. Both of the two outer gap models, established by \citet{Aleksic_Gap_2012} and \citet{Harding_Gap_2015} respectively, fail to describe the spectral shape observed in 1--10~GeV. The sum of the freshly-accelerated wind's IC emission modelled by \citet{Aharonian_Wind_2012} and the extrapolation of our PLSEC model can account for the $\gamma$-ray spectrum up to $\sim$400~GeV. This implies that our PLSEC can be interpreted as a nominal component (i.e., synchro-curvature radiation from the outer gap and/or wind). On the other hand, \citet{Mochol_Wind_2015} established a model for a current sheet of striped wind which can, in absence of an outer-gap component, satisfactorily reproduce the observed flux and spectral shape from 1~GeV to 1~TeV. In this model, the transition from the synchrotron-dominated spectrum to the SSC-dominated spectrum occurs at $\sim$300~GeV.

All in all, we propose a hybrid scenario where different acceleration sites account for pulsed $\gamma$-rays of the Crab pulsar at different phases and energies. \DHA{Roughly speaking, the polar cap is responsible for the emission at and around the bridge below 10~GeV, the outer gap is responsible for $<$10~GeV emissions at other phases, and the wind is responsible for  emissions above 10~GeV at any phase.} In a more detailed modelling approach, one should carefully deal with the transition phases and transition energies among different emission components.

\begin{acknowledgements}
PKHY acknowledges the support of the DFG under the research grant 	HO 3305/4-1. PKHY gives sincere gratitude to D. Horns for useful discussions and for his encouragement regarding my submission as a single author. We greatly appreciate M. Kerr for providing the ephemeris of the Crab pulsar for phased analysis. We thank D. Carreto Fidalgo for his help in reconstructing Figure~\ref{ShowCase}. We thank the anonymous referee for very useful comments which helped to improve the manuscript.
\end{acknowledgements}

\vspace{5mm}
%%%\facilities{}
%%%\software{Fermi Science Tools (v11r5p3) and DELightcurveSimulation \citep{Connolly2015, Emmanoulopoulos2013}}

\bibliographystyle{aa}
\InputIfFileExists{Crab_pulsar_paper.bbl}



\end{document}

\title{Inferring the origins of the pulsed $\gamma$-ray emission from the Crab pulsar with ten-year \emph{Fermi} LAT data \\ (Corrigendum)}

\titlerunning{Erratum for: A\&A 640, A43 (2020)}

   \author{Paul K. H. Yeung
          \inst{1,2}
          }

   \institute{Institute of Experimental Physics, Department of Physics, University of Hamburg, Luruper Chaussee 149, D-22761 Hamburg, Germany\\
             \email{kin.hang.yeung@desy.de}
\and
Nicolaus Copernicus Astronomical Center, Polish Academy of Sciences, Rabia\'nska 8, 87-100, Toru\'n, Poland
}

\abstract
{A\&A 640, A43 (2020), \url{https://doi.org/10.1051/0004-6361/202038166}}

% \abstract{}{}{}{}{} 
% 5 {} token are mandatory


   \maketitle
%
%-------------------------------------------------------------------

This is a corrigendum to the original article \citet{yeung_pulsar_2020} and is mostly adapted from Appendix C.4 of the author's PhD thesis \citep{yeung2021studies}.

\begin{figure}
	\includegraphics[width=9cm]{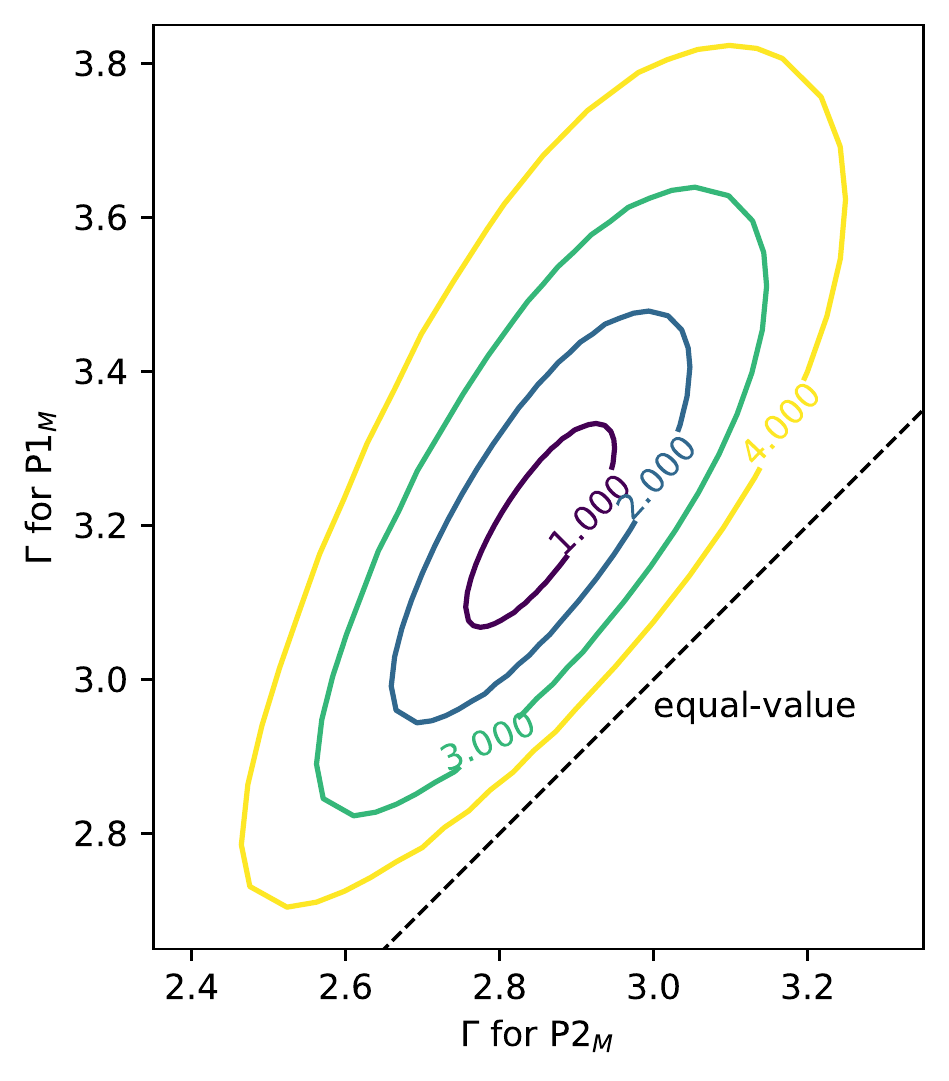}
	
	\caption{Confidence regions of $\Gamma$ values for P1$_M$ and P2$_M$. The plotted covariance contours respectively correspond to the significance levels of $1\sigma$, $2\sigma$, $3\sigma$, and $4\sigma$ (from innermost to outermost). The reference line (which is black, dashed, and straight) indicates the combinations of equal $\Gamma$ values.   \label{mncontour}}
\end{figure}

In Section 3.4 of the original article, we deduced from the ``power law with a scaling factor" (PLSF) fitting result that $\Gamma$ for P2$_M$ is lower than that for P1$_M$. Nevertheless, the previous calculation of the significance of this difference did not consider the covariance of the $\Gamma$ values for P1$_M$ and P2$_M$ (be noted that the spectral data for P1$_M$ and P2$_M$ are fit simultaneously such that their solutions share the same scaling factor $\epsilon$). After taking  this covariance into account,  which is computed to be 0.010, we now determine that $\Gamma$ for P2$_M$ is  lower than that for P1$_M$ by $0.35\pm0.08$, which corresponds to $\sim4.2\sigma$ (much higher than the $\sim2.1\sigma$ reported in the original paper). Additionally, Figure~\ref{mncontour} demonstrates that the $4\sigma$ confidence region of $\Gamma$ values for P1$_M$ and P2$_M$ excludes all equal-value combinations. 

The revised significance strengthens the argument that the flux ratio of P1 to P2 still decreases with photon energy even above 10 GeV. This, in turn,  strengthens the interpretation that  there are anisotropic pulsar winds that could lead to the observed harder spectrum at P2$_M$ relative to P1$_M$ \citep[as put forward by][]{Aharonian_Wind_2012}.

Moreover, taking the statistical errors into account,  the $\epsilon$ values obtained for MAGIC and VERITAS (both $\sim$1.22) are basically consistent with the estimated systematic errors of around 15\% on energy scales of IACT telescopes under  perfect atmospheric conditions \citep[e.g.,][]{Aharonian_IACT_2006, Aleksic_IACT_2016}. This provides further confirmation for the physical reasonableness of those fitting results.